\documentstyle[prl,aps,psfig,preprint]{revtex}
\draft
\global\firstfigfalse
\tightenlines
\newcommand{\nc}{\newcommand}\newcommand{\rnc}{\renewcommand}
\rnc{\/}{\over}\nc{\<}{\left\langle}\rnc{\>}{\right\rangle}\rnc{\/}{\over}
\rnc{\(}{\left (}\rnc{\)}{\right )}\rnc{\[}{\left [}\rnc{\]}{\right ]}
\nc{\Tr}{\text{tr}}\rnc{\i}{\text{i}}\nc{\e}{\text{e}}
\nc{\nq}{{\cal N}}
\rnc{\mark}[1]{{\bf #1}}
\nc{\no}[1]{}
\begin{document}
\title{Combinatorial Identities from the Spectral Theory of Quantum Graphs}
\author{Holger Schanz$^{\dag}$ and Uzy Smilansky$^{\ddag}$\\[5mm]}
\address {$^{\dag}$Universit{\"a}t G{\"o}ttingen and
MPI f{\"u}r Str{\"o}mungsforschung,\\ 37073 G\"ottingen, Germany\\[2mm]}
\address{$^{\ddag}${Department of Physics of Complex Systems,} \\
{The Weizmann Institute of Science, Rehovot 76100, Israel}\\[5mm]}
\date{March 27, 2000}
\maketitle
%%%%%%%%%%%%%%%%%%%%%%%%%%%%%%%%%%%%%%%%%%%%%%%%%%%%%%%%%%%%%%%%%%%%%%%%%%%%%%%%
\begin{center}\mbox{}\\
submitted to the\\
{\it  Special volume of The Electronic Journal of Combinatorics honoring\\
Professor Aviezri Fraenkel}\\
\end{center}
\begin{abstract}
We present a few combinatorial identities which were encountered in our work
on the spectral theory of quantum graphs. They establish a new connection
between the theory of random matrix ensembles and combinatorics.
\end{abstract}

%-------------------------------------------------------------------------
\section {{\bf Introduction} \label{introduction}}
\noindent 
In the present paper we show that some questions arising in the study of
spectral correlations for quantum graphs can be cast as combinatorial
problems. Solving these problems for a particular system, we discovered the
following novel combinatorial identities:
\begin{description}
\item{(i)} Let $n, q$ be arbitrary integers with $1\le q <n$ and
$$
F_{\nu,\nu'}(n,q)={(n-1)n\/2}{(-1)^{\nu+\nu'} \over \nu\nu'}
{n\choose \nu+\nu'}^{-1}
{q-1\choose
\nu-1}{q-1\choose \nu'-1}{n-q-1\choose\nu-1}{n-q-1\choose\nu'-1}\,.
$$
Then
\begin{equation}\label{ci3}
S(n,q)=\sum_{\nu,\nu'=1}^{\min(q,n-q)}F_{\nu,\nu'}(n,q)=1\,.
\end{equation}
\item{(ii)} Let $s,t$ be arbitrary positive integers and
\begin{equation}\label{nq}
\nq(s,t)=\sum_{\nu=1}^{\min(s,t)}(-1)^{t-\nu}{t\choose\nu}{s-1\choose \nu-1}
=(-1)^{s+t}{s+t-1\choose s}^{1/2}P_{s+t-1,s}(t)\ ,
\end{equation}
where $P_{N,k}(x)$ are the Kravtchouk polynomials to be defined in
Eq.~(\ref{kravtchouk}). Further, let $x,y$ be complex with
$|x|,|y|<1/\sqrt{2}$.  Then we have the generating functions
\begin{equation}\label{g1def}
G_{1}(x)=
\sum_{s,t=1}^{\infty}\nq^{2}(s,t)\,x^{s+t}
={x\over 2x-1}\({1\over \sqrt{4x^2+1}}-{1\/1-x}\)\,,
\end{equation}
\begin{equation}\label{g2def}
G_{2}(x)=
\sum_{s,t=1}^{\infty}\nq(s,t)\,\nq(t,s)\,x^{s+t}
={1\over 2}{4x^2+2x+1\/(2x+1)\sqrt{4x^2+1}}-{1\over 2}
\end{equation}
and
\begin{equation}\label{gdef}
g(x,y)=
\sum_{s,t=1}^{\infty}\nq(s,t)\,x^{s}\,y^{t}{xy\over(1+y)(1-x+y-2xy)}\,.
\end{equation}
\item{(iii)} Let $m$ be any positive integer. Then
\begin{equation}\label{ci2a}
4m^{2}\sum_{q=1}^{2m-1}\({\nq(s,t)\over
q}\)^{2}=2^{2m+1}+(-1)^{m}{2m\choose m}-2
\end{equation}
and
\begin{equation}\label{ci2b}
(2m+1)^{2}\sum_{q=1}^{2m}\({\nq(s,t)\over
q}\)^{2}=2^{2m+2}-2\,(-1)^{m}{2m\choose m}-2\,.
\end{equation}
\item{(iv)}
Let $0\le q \le n$, and define
\begin{equation}\label{adef}
A(n,q) =  {1\over\sqrt{2^{n}}} \left \{
\begin{array}{ll}
  1& 
 {\rm for}\;q=0,n  \nonumber \\
  (-1)^{q}(n/q)\,\nq(n-q,q)\;\;\;&{\rm for}\;0<q<n \ . \\
\end{array}
\right .
\end{equation}
Then for any positive integers $0 \le \kappa \le \nu $  and an arbitrary
integer $n_0$,
\begin {equation}
\lim_{\epsilon \rightarrow 0} \ \epsilon \ \sum _{n=n_0}^{\infty}
{\rm e}^{-n\epsilon} \sum _{q=0}^n A(n+\nu , q+\kappa) A(n,q)   = A(\nu,\kappa)
\ .
\label{traceid}
\end{equation}

\end{description}
These identities establish a new connection between combinatorics and the
theory of random ensembles of $2\times 2$ matrices. The physical background
and motivations are described in a few recent publications \cite{KS97,SS00},
to which the interested reader is referred. An immediate application of
(\ref{ci3}) is also given in \cite{SS00}. Here, we shall provide the minimum
background necessary for the understanding of the combinatorial aspects of the
problem, and for a self-contained exposition of our results. It will also
enable us to propose a conjecture which generalizes the combinatorial approach
to random matrix theory for matrices of large dimensions.

 We start with a few definitions: Graphs consist of $V$ {\it vertices}
connected by $B$ {\it bonds} (or {\it edges}). The {\it valency} $v_{i}$ of a
vertex $i$ is the number of bonds meeting at that vertex. We shall assume that
two vertices can be connected by a single bond at most. We denote the bonds
connecting the vertices $i$ and $j$ by $b=[i,j]$.  The notation $[i,j]$ will
be used whenever we do not need to specify the {\it direction} on the bond.
Hence $[i,j]=[j,i]$.  {\it Directed bonds} will be denoted by $d=(i,j)$, and
we shall always use the convention that the bond is directed from the first to
the second index. If $d=(i,j)$ we use the notation $\hat d=(j,i)$. Let
$g^{(i)}$ be the set of directed bonds $(i,j)$ which emanate from the vertex
$i$, and $\hat g^{(i)}$ the set of directed bonds $(j',i)$ which converge at
$i$. The vertices $i$ and $j$ are connected if $ g^{(i)}\cap \hat g^{(j)} \ne
\emptyset$. The bond $d'$ is {\it connected} to the bond $d$ if there is some
vertex $i$ with $d \in g^{(i)}$ and $d'\in\hat g^{(i)}$.
$d$ and $\hat d$ are always connected.

 The Schr\"odinger operator on the graph is defined after the natural metric
is assigned  to the bonds, and the solutions of the one-dimensional
Schr\"odinger equation on each bond $(\i\,{\rm d}_x -A)^2 \psi(x) =k^2\psi(x)$
is given as a linear combination of counter-propagating waves. $A$ stands here
for a magnetic flux.  The (complex) amplitudes of the counter propagating waves
are denoted by $a_{d}$, where the subscript $d$ stands for the directed bond
along which the wave propagates, $d=1,..,2B$. Appropriate boundary conditions
at the vertices are imposed, and the spectrum of the Schr\"odinger operator on
the graph is determined as the (infinite, discrete) set of energies $k_n^2$,
for which there exists a non-trivial set of $a_d$ which is consistent with the
boundary conditions. The condition of consistency can be expressed by the
requirement that
\begin{equation}
\det(I-S(k))=0
\label{secular}
\end{equation}
where the {\em bond-scattering matrix} $S(k)$ is a unitary operator in the
Hilbert space of $2B$ dimensional vectors of coefficients $a_d$. The unitarity
of $S(k)$ ensures that the spectrum of the Schr\"odinger operator is real. The
matrix $S(k)$, which is the object of our study is defined as
\begin{equation}
{S}_{(i,j),(l,m)}(k) = \delta _{j,l}
{\rm e}^{\i\phi_{(i,j)}(k)}\sigma_{i,m}^{(j)}(k)\,.
\label{Sdef}
\end{equation}
The matrix elements $S_{d,d'}(k)$ vanish if the bonds are not connected. As a
consequence, the unitarity of $S$ implies also the unitarity of the
$v_{j}$-dimensional {\em vertex-scattering matrices} $\sigma_{i,m}^{(j)}$ and
vice versa.  The phases $\phi_{(i,j)}$, are given in terms of the bond length
$L_{[i,j]}$, and the magnetic flux $A_{(i,j)}=-A_{(j,i)}$ \cite{KS97},
\begin{equation}
\label{phases}
\phi_{(i,j)}(k)=(k+A_{(i,j)})\,L_{[i,j]} \ .
\end{equation}
The two phases pertaining to the same bond $\phi_{d}$ and $\phi_{\hat d}$ are
equal when $A_{d}=A_{\hat d}=0$. In this case $S$ is symmetric and the
Schr\"odinger operator on the graph is invariant under {\em time
reversal}. Time-reversal symmetry is violated when some magnetic fluxes do not
vanish.

$S$ can also be interpreted as a quantum time evolution operator describing
the scattering of waves with wave number $k$ between connected bonds. The wave
gains the phase $\phi_{(i,j)}(k)$ during the propagation along the bond
$(i,j)$, while the $\sigma_{i,m}^{(j)}$ describe the scattering at the
vertices.  In this picture, the unitarity of $S$ guarantees the conservation
of probability during the time evolution.

We will avoid unnecessary technical difficulties and consider the matrices
$\sigma_{i,m}^{(j)}$ to be $k$ independent constants.  One may find explicit
expressions for $\sigma_{i,m}^{(j)}$ by requiring besides unitarity that the
wave function is continuous at the vertices.  The resulting expression is
\cite{KS97}
\begin{equation}
\sigma _{j,j^{\prime }}^{(i)}={2\over v_{i}}-\delta _{j,j^{\prime }}
\qquad\mbox{(Neumann b.~c.)}\,.
\label{signeu}
\end{equation}
Note that back-scattering ($j =j'$) is singled out both in sign and in
magnitude. In all nontrivial cases ($v_{i}>2$) the back-scattering amplitude
is negative, and the back-scattering probability $|\sigma_{j,j}^{(i)}|^2$
approaches $1$ as the valency $v_i$ increases.

 Finally, a ``classical analogue" of the quantum dynamics can be defined as a
random walk on the directed bonds, in which the transition probability between
bonds $(i,j)$, $(j,l)$ connected at vertex $j$ is
$|\sigma_{i,l}^{(j)}|^2$. The resulting classical evolution operator with
matrix elements
\begin{equation}
\label{ctevol}
U_{d,d'}=|S_{d,d'}|^{2}
\end{equation}
is probability conserving, since unitarity implies $\sum_i |\sigma
_{i,l}^{(j)}|^2=1$.

\section{Spectral Two-Point Correlations and Periodic-Orbit Sums}

The spectrum of $S$ consists of $2B$ eigenvalues ${\rm e}^{i\theta_l}$ which
are confined to the unit circle. Their distribution is given in terms of the
spectral density
\begin{equation}
d(\theta)\equiv \sum_{l=1}^{2B}\delta_{2\pi}(\theta-\theta_l)=\frac{2B}{2\pi}+
\frac 1{2\pi}\sum_{n=1}^\infty s_{n}\e^{-\i\theta n} +{\rm c.c.}\,,
\label{sms1}
\end{equation}
where $\delta_{2\pi}$ denotes the $2\pi$ periodic delta function. The first
term on the r.h.s.\ is the average density $\overline{d}=\frac{2B}{2\pi}$. The
coefficients of the oscillatory part
$$s_{n}={\rm tr}S^n$$ will play an important r\^ole in the following.
$s_n={\rm tr}S^n$ is a sum over products of $n$ matrix elements of $S$, and
because of (\ref{Sdef}) the bond indices of each summand describe a connected
$n$-cycle ($\equiv$ $n$-periodic orbit) on the graph
\begin{equation}
s_n=\sum_{p\in {\cal P}_n}{\cal A}_p\,{\rm e}^{i\Phi_{p}}\,.
\label{posum}
\end{equation}
In (\ref{posum}) ${\cal P}_n$ denotes the set of all $n$-{\em periodic orbits}
(PO's) on the graph. Note that for the convenience of presentation we will
consider cycles differing only by a cyclic permutation as different PO's.  The
phases $\Phi_p=\sum_{j=0}^{n-1}\phi_{d_{j}}$ can be interpreted as the action
along the PO $p$. The amplitudes ${\cal A}_p$ are given by
\begin{equation}
{\cal A}_p=\prod _{j=0}^{n-1} S_{d_{j+1},d_{j}}\,,
\end{equation}
where $j$ is understood mod $n$. Sometimes it is useful to split the amplitude
in its absolute value and a phase ${\rm e}^{\i\mu_{p}\pi}$.  For example, in
the case of Neumann b.~c. (\ref{signeu}) $\mu_{p}$ is an integer counting the
number of back-scatterings along $p$.

In complete analogy to (\ref{posum}) we can represent also the traces of
powers of the classical evolution operator
\begin{equation}
\label{classtr}
u_{n}={\rm tr}U^{n}
\end{equation}
as sums of periodic orbits of the graph.

The two-point correlations in the spectrum of $S$ (\ref{sms1}) can be
expressed in terms of the average excess probability density $R_2(r;\beta)$ of
finding two phases at a distance $r$, where $r$ is measured in units of the
mean spacing $\frac{2\pi}{2B}$,
\begin{eqnarray}
R_2(r;\beta)&=&
\<{1\over 2B}\int_{-\pi}^{+\pi}{\rm d}\theta\;
d\(\theta\)\,d\(\theta-[\pi/B]\,r\)\>_{\beta}
-{B\over \pi}
\nonumber\\ &=&
{2\over 2\pi}\sum_{n=1}^\infty \cos\(\frac{\pi}{B}nr\)
\frac 1{2B}\<|s_{n}|^2\>_{\beta}\,.
\label{sms3}
\end{eqnarray}
The bond scattering matrix depends parametrically on the phases $\phi_{d}$
(\ref{phases}). We shall define two statistical ensembles for $S$ in the
following way.  The ensemble for which time-reversal symmetry is broken
consists of $S$ matrices for which the $\phi_{d}$ are all different, and we
consider them as independent variables distributed uniformly on the $2B$
torus. Invariance under time reversal implies $\phi_{d}=\phi_{\hat d}$ and the
corresponding ensemble is defined in terms of $B$ independent and uniformly
distributed phases.  We shall distinguish between these ensembles by the value
of the parameter $\beta = \{$number of independent phases$\}/B$.
Expectation values with respect to these measures are denoted in (\ref{sms3})
by triangular brackets,
\begin{equation}
\label{average}
\<\dots\>_{\beta} \equiv \prod_{d}^{\beta B}\(\frac {1}{2\pi}
\int_{-\pi}^{+\pi}{\rm d}\phi_{d}\)\,
\dots\,.
\end{equation}
The Fourier transform of  $R_2(r;\beta)$ is the {\em form factor}
\begin{equation}\label{ff}
K(n/2B;{\beta})={\frac 1{2B}}\<|s_{n}|^2\>_{\beta}
\end{equation}
on which our interest will be focussed.  If the eigenvalues of the $S$ were
statistically independent and uniformly distributed on the unit circle,
$K(n/2B)=1$ for all $n$.  Any deviation of the form factor from unity implies
spectral correlations. Using (\ref{posum}) the form factor (\ref{ff}) is
expressed as a double sum over PO's
\begin{eqnarray}
K(n/2B;\beta)&=&\frac 1{2B}\<\left|\sum_{p\in {\cal P}_n}{\cal A}_p
\e^{\i\Phi_{p}}\right|^2 \>_{\beta}
\label{sms5}
\\&=&
\frac 1{2B}\sum_{p,p'\in {\cal P}_n}{\cal A}_p{\cal A}_{p\prime}^{*}
\<\e^{\i(\Phi_{p}-\Phi_{p'})}\>_{\beta}
\nonumber
\end{eqnarray}
In order to perform the average over all the phases $\phi_{d}$ in (\ref{sms5})
we write
\begin{eqnarray}
\Phi_{p}=\sum_{d}n^{(p)}_{d}\,\phi_{d}\,,
\end{eqnarray}
where $n_{d}^{(p)}$ counts the number of traversals of each directed bond such
that $\sum_{d}n_{d}^{(p)}=n$. According to (\ref{average}) we have
\begin{eqnarray}
\label{without_trs}
\<\e^{\i\,(n\phi_{d}+n'\phi_{d'})}\>_{\beta=1}&=&\delta_{n,0}\delta_{n',0}\,,
\\
\label{with_trs}
\<\e^{\i\,(n\phi_{d}+\hat n\phi_{\hat d})}\>_{\beta=2}&=&\delta_{n+\hat
n,0} \,.
\end{eqnarray}
Thus, the double sum in (\ref{sms5}) can be restricted to families of orbits.
For $\beta =2$, let ${\cal L}$ be the {\em family of isometric PO's} which
have the same integers $n_{d}^{({\cal L})}$. That is, the family consists of
all the PO's which traverse the same directed bonds the same number of times,
but not necessarily in the same order. In the case $\beta =1$, ${\cal L}$
consists of all PO's sharing $n_b^{({\cal L})} \equiv n_{d}^{({\cal
L})}+n_{\hat d}^{({\cal L})}$. That is, the family contains all PO's which
traverse the same set of undirected bonds the same number of times,
irrespective of direction or order.  We find
\begin{eqnarray}
K(n/2B;\beta )&=&
\frac 1{2B}\sum_{{\cal L}\in {\cal F(\beta)}_n}
|\sum_{p\in{\cal L}}{\cal A}_p|^{2}\,.
\label{famsum}
\end{eqnarray}
${\cal F(\beta)}_{n}$ denotes the set of all vectors ${\cal L}=[n_{d}]$ for
$\beta=2$ (${\cal L}=[n_{b}]$ for $\beta=1$) of $\beta\,B$ non-negative
integers summing to $n$, for which at least one PO exists.  For Neumann
b.~c. (\ref{signeu}), e.~g., (\ref{famsum}) amounts to counting the PO's in a
given set ${\cal L}=[n_{d}]$ taking into account the number of
back-scatterings along the orbit. The problem of spectral statistics is now
reduced to a counting (combinatorial) problem which is, however, very
complicated in general.  Even the determination of the number of families
${\cal L}$ for a given $n$ is difficult.  For $\beta=2$ an obvious necessary
condition for the existence of a PO with a given set of bond traversals ${\cal
L}=[n_{d}]$ is that at any vertex the number of incoming and outgoing bonds is
the same, i.~e.\
$$
\sum_{d\in g^{(i)}}n_{d}=\sum_{d\in \hat g^{(i)}}n_{d}
\qquad
(i=1,\dots V)\,.
$$
For $\beta=1$, the analogous  condition reads
$$
\sum_{d\in g^{(i)}}n_{d}-\sum_{d\in \hat g^{(i)}}n_{d} \mbox{ mod }2=0
\qquad
(i=1,\dots V)\,,
$$
i.~e.\ the total number of traversals of adjacent bonds should be even at each
vertex. However, it is not so easy to formulate a sufficient condition for the
existence of a PO given a set of numbers $n_{d}$. In particular one must take
care to exclude cases, in which the set of traversed bonds is a union of two
or more disconnected groups (``composite orbits").

Extensive numerical work \cite {KS97} revealed that for fully connected graphs
($v_{j}\equiv V-1$), and for $V \gg 1$, the form-factor (\ref{ff}) is well
reproduced by the predictions of random matrix theory \cite {Dyson} for the
Circular Orthogonal Ensemble (COE) ($\beta =1$) or the Circular Unitary
Ensemble (CUE) ($\beta =2$). This leads us to expect that (\ref{famsum})
approaches the corresponding random matrix prediction in the limit
$V\rightarrow \infty$. This conjecture is proposed as a challenge to
asymptotic combinatorial theory.

%%%%%%%%%%%%%%%%%%%%%%%%%%%%%%%%%%%%%%%%%%%%%%%%%%%%%%%%%%%%%%%%%%%%%%%%%%%
\section{\bf The Ring Graph}

In the following we will evaluate explicitly the quantities introduced in the
previous section for one of the simplest quantum graphs. It consists of a
single vertex on a loop (see fig.~\ref{tst}). There are two directed bonds
$d=1$ and $\hat d=2$ with $\phi_{1}\ne \phi_{2}$, i.~e.\ time-reversal
symmetry is broken. Since this graph would be trivial for Neumann b.c. the
vertex-scattering matrix at the only vertex is chosen as
\begin {equation}
\sigma(\eta) =
\({\begin{array} {ll}
\cos \eta & {\rm i}\sin \eta \\
{\rm i}\sin\eta  & \cos\eta
\end{array}}  \)\,,
\label {2-starsigma}
\end{equation}
with $0\le\eta\le\pi/2$. The corresponding bond-scattering matrix is
\begin {equation}
S(\eta) =
\({\begin{array} {ll}
{\rm e}^{\phi_1} & 0   \\
0 &  {\rm e}^{\phi_2}
\end{array}}\)
\(
{\begin{array} {ll}
\cos\eta & {\rm i}\sin \eta \\
{\rm i}\sin\eta  & \cos\eta
\end{array}}\)\,.
\label{2-starSB}
\end{equation}
We shall compute the form factor for two ensembles. The first is defined by a
fixed value of $\eta=\pi/4$, and the only average is over the phases
$\phi_{d}$ according to (\ref{without_trs}). The second ensemble includes an
additional averaging over the parameter $\eta$. We will show that the measure
for the integration over $\eta$ can be chosen such that the model yields
exactly the CUE form factor for 2$\times$ 2 random matrices \cite{Dyson}.
%%%%%%%%%%%%%%%%%%%%%%%%%%%%%%%%%%%%%%%%%%%%%%%%%%%%%%%%%%%%%%%%%%%%%%%%%%%%%%%%
\subsection {\bf Periodic Orbit Representation of $u_n$}
We will first illustrate our method of deriving combinatorial results from the
ring graph in a case where a known identity is obtained.  Consider the
classical evolution operator $U$ of the ring graph. According to
(\ref{ctevol}) we have
\begin{equation} U(\eta)=\(
{\begin{array} {ll} \cos ^2\eta & \sin ^2 \eta \\
\sin ^2\eta & \cos ^2 \eta
\end{array}}  \right )\,.
\label{2-starclass}
\end{equation}
The spectrum of $U$ consists of $\{1,\cos 2\eta \}$, such that
\begin{equation}\label{un}
u_n (\eta )=1+\cos ^n 2\eta\,.
\end{equation}
We will now show how this result can be obtained from a sum over the periodic
orbits of the system, grouped into families of orbits as in (\ref{famsum}). In
the classical calculation it is actually not necessary to take the families
into account, but we would like to stress the analogy to the quantum case
considered below.  The periodic orbit expansion of the classical return
probability can easily be obtained from (\ref{2-starclass}) by expanding all
matrix products in (\ref{classtr}). We find
\begin{eqnarray}\label{un_po}
u_{n}&=&
\sum_{i_{1}=1,2}
\dots
\sum_{i_{n}=1,2}\prod_{j=0}^{n-1}U_{i_{j},i_{j+1}}(\eta)\,,
\end{eqnarray}
where $j$ is again taken mod~$n$. In the following the binary sequence
$[i_{j}]$ ($i_{j}\in\{1,2\}$; $j=0,\dots,n-1$) is referred to as the code of
the orbit.  We will now sort the terms in the multiple sum above into families
of isometric orbits. In the present case a family is completely specified by
the integer $q\equiv q_1$ which counts the traversals of the loop $1$, i.e.,
the number of letters $1$ in the code word. Each of these $q$ letters is
followed by an uninterrupted sequence of $t_{j}\ge 0$ letters $2$ with the
restriction that the total number of letters $2$ is given by
\begin{equation}
\sum_{j=1}^{q}t_{j}=n-q\,.
\end{equation}
We conclude that each code word in a family $0<q<n$ which starts with
$i_{1}=1$ corresponds to an ordered partition of the number $n-q$ into $q$
non-negative integers, while the words starting with $i_{1}=2$ can be viewed
as partition of $q$ into $n-q$ summands.

To make this step very clear, consider the following example: All code words
of length $n=5$ in the family $q=2$ are $11222$, $12122$, $12212$, $12221$ and
$22211$, $22121$, $21221$, $22112$, $21212$, $21122$. The first four words
correspond to the partitions $0+3=1+2=2+1=3+0$ of $n-q=3$ into $q=2$ terms,
while the remaining $5$ words correspond to
$2=0+0+2=0+1+1=1+0+1=0+2+0=1+1+0=2+0+0$.

In the multiple products in (\ref{un_po}) a backward scattering along the
orbit is expressed by two different consecutive symbols $i_{j}\ne i_{j+1}$ in
the code and leads to a factor $\sin^2\eta$, while a forward scattering
contributes a factor $\cos^2\eta$ . Since the sum is over periodic orbits, the
number of back scatterings is always even and we denote it with $2\nu$. It is
then easy to see that $\nu$ corresponds to the number of positive terms in the
partitions introduced above, since each such term corresponds to an
uninterrupted sequence of symbols $2$ enclosed between two symbols $1$ or vice
versa and thus contributes two back scatterings. For the codes starting
with a symbol $1$ there are ${q\choose \nu}$ ways to choose the $\nu$ positive
terms in the sum of $q$ terms, and there are ${n-q-1\choose \nu-1}$ ways to
decompose $n-q$ into $\nu$ {\em positive} summands. After similar reasoning
for the codes starting with the symbol $2$ we find for the periodic orbit
expansion of the classical return probability
\begin{eqnarray}
u_{n}(\eta)&=&2\cos^{2n}\eta+\sum_{q=1
}^{n-1}\sum_{\nu}
\[{q\choose \nu}{n-q-1\choose \nu-1}+{n-q\choose
\nu}{q-1\choose \nu-1}\]
\sin^{4\nu}\!\eta\,\cos^{2n-4\nu}\!\eta
\nonumber
\\
&=&2\cos^{2n}\eta+\sum_{q=1}^{n-1}\sum_{\nu}{n\/\nu}
{q-1\choose\nu-1}{n-q-1\choose\nu-1}\sin^{4\nu}\!\eta\,\cos^{2n-4\nu}\!\eta\,
\label{ringcsum}
\end{eqnarray}
The summation limits for the variable $\nu$ are implicit since all terms
outside vanish due to the properties of the binomial coefficients. From the
equivalence between (\ref{un}) and (\ref{ringcsum}) the combinatorial identity
\begin{equation}\label{ci1}
\sum_{q=1}^{n-1}{q-1\choose \nu-1}{n-q-1\choose \nu-1}=
{n-1\choose 2\nu-1}={2\nu\over n}{n\choose 2\nu}\,.
\end{equation}
could be deduced which indeed reduces (\ref{ringcsum}) to a form
\begin{eqnarray}
u_{n}(\eta)&=&2\sum_{\nu}{n\choose
2\nu}\sin^{4\nu}\!\eta\,\cos^{2n-4\nu}\!\eta
\nonumber
\\
&=&(\cos^{2}\!\eta+\sin^{2}\!\eta)^{n}+(\cos^{2}\!\eta-\sin^{2}\!\eta)^{n}\,,
\label{ringcsumex}
\end{eqnarray}
which is obviously equivalent to (\ref{un}). 

(\ref{ci1}) can also be derived by some straight forward variable
substitutions from the identity
\begin{equation}
\sum_{k=l}^{n-m}{k\choose l}{n-k\choose m}={n+1\choose l+m+1}
\end{equation}
which is found in the literature \cite{prudnikov}.
%%%%%%%%%%%%%%%%%%%%%%%%%%%%%%%%%%%%%%%%%%%%%%%%%%%%%%%%%%%%%%%%%%%%%%%%%%%%%%%%
\subsection {\bf Quantum Mechanics: Spacing Distribution and Form
Factor}\label{qm}
In the following two subsections we derive novel combinatorial identities
by applying the reasoning which led to (\ref{ci1}) to the quantum evolution
operator (\ref {2-starSB}) of the ring graph. We can write the eigenvalues of
$S(\eta)$ as ${\rm e}^{\i(\phi_{1}+\phi_{2})/2}\,{\rm e}^{\pm \i\lambda/2}$
with
\begin{equation}
\lambda=2\,{\rm arcos}\[\cos\eta\,\cos \({\phi_1-\phi_2\/2}\)\]
\label
{2-starlambda}
\end{equation}
denoting the difference between the eigenphases. For the two-point correlator
we find
\begin{eqnarray}
R_2(r,\eta)&=&
\<{1\over 2}\int_{-\pi}^{+\pi}{\rm d}\theta\;
d\(\theta+{\pi\,r\/2}\)\,d\(\theta-{\pi\,r\/2}\)\>_{\phi_{1,2}}-{1\over \pi}
\nonumber\\&=&
{\delta_{2}(r)-1\over \pi}
\<{\delta_{2\pi}\(\pi\,r+\lambda\)+
   \delta_{2\pi}\(\pi\,r-\lambda\)\/2}\>_{\phi_{1,2}}
\nonumber\\&=&
{\delta_{2}\(r\)-1\over \pi}+
{\sin|\pi\,r/2|\over 2\pi}
{\Theta(\cos^{2}\eta-\cos^{2}(\pi\,r/2))\over
\sqrt{\cos^{2}\eta-\cos^{2}(\pi\,r/2)}}
\label{r2eta})
\end{eqnarray}
Here, $\delta_{2}\(r\)$ is the $2$-periodic $\delta $ function.  In particular
for equal transmission and reflection probability ($\eta=\pi/4$) we have
\begin{eqnarray}
R_2(r,\pi/4)&=&{\delta_{2}\(r\)-1\/\pi}+
{1\over 2\pi}\sqrt{{\cos(\pi r)-1\/\cos(\pi r)}}
\Theta\({1\/2}-|r-1|\)
\end{eqnarray}
and, by a Fourier transformation, we can compute the form factor
\begin{eqnarray}
K(n,\pi/4)&=&\pi\int_{0}^{2}{\rm d}r\,\cos\(n\pi r\)\,R_{2}(r,\pi/4)
\nonumber\\&=&
1+{(-1)^{m+n}\over 2^{2m+1}}{2m\choose m}-{3\over 2}\delta_{n,0}
\label{K2PI4_UZY}
\\&\approx&
1+{(-1)^{m+n}\over 2\sqrt{\pi n}}\qquad(n\gg 1)\, ,
\label{k2pi4_uzy_app}
\end{eqnarray}
where $ m =[n/2]$ and $[\cdot]$ stands for the integer part.

Next we consider the ensemble for which transmission and reflection
probabilities are uniformly distributed between $0$ and $1$. For the parameter
$\eta$ this corresponds to the measure ${\rm d}\mu(\eta)=2|\cos\eta \sin\eta
|{\rm d}\eta$. The main reason for this choice is that upon integrating
(\ref{r2eta}) one gets
\begin{eqnarray}
R_2^{\rm (av)}(r)&=&
{\delta_{2}\(r\)-1\over \pi}+{\sin^{2}(\pi\,r/2)\over \pi}
\end{eqnarray}
which coincides with the CUE result for $2\times 2$ matrices. A Fourier
transformation results in
\begin {equation}
K_2(n)=\left\{{
 \begin{array}{ll}
 {1\over 2} & {\rm for}\;n=1 \\
 1 & {\rm for}\;n\ge 2
 \end{array} }
\right.\,.
\label {2-starK(n)CUE}
\end{equation}
The form factors (\ref{K2PI4_UZY}), (\ref{k2pi4_uzy_app}) and
(\ref{2-starK(n)CUE}) are displayed in Fig.~\ref{tst} below.
%%%%%%%%%%%%%%%%%%%%%%%%%%%%%%%%%%%%%%%%%%%%%%%%%%%%%%%%%%%%%%%%%%%%%%%%%%%%%%%%
\subsection{Periodic Orbit Expansion of the Form Factor}\label{po}
An explicit formulation of (\ref{famsum}) for the ring graph is found by
labelling and grouping orbits as explained in the derivation of
(\ref{ringcsum}). We obtain
\begin{eqnarray}\label{K2eta}
K_2(n;\eta)&=&\cos^{2n}\!\eta+{n^2\over 2}
\sum_{q=1}^{n-1}
\[\sum_{\nu}{(-1)^{\nu}\/\nu}{q-1\choose\nu-1}{n-q-1\choose\nu-1}
\sin^{2\nu}\!\eta\,\cos^{n-2\nu}\!\eta\]^{2}\,,
\end{eqnarray}
where $q$ denotes the number of traversals of the ring in positive direction
and $2\nu$ is the number of backward scatterings along the orbit. The inner
sum over $\nu$ can be written in terms of Kravtchouk polynomials as
\begin{eqnarray}\label{K2eta_K}
K_2(n;\eta)&=&
\cos^{2n}\!\eta+{1\/2}\sum_{q=1}^{n-1}
{n-1\choose
n-q}\cos^{2q}\!\eta\sin^{2(n-q)}\!\eta
\[{n\over
q}P_{n-1,n-q}^{(\cos^{2}\!\eta,\sin^2\!\eta)}(q)\]^2\,,
\end{eqnarray}
and the Kravtchouk polynomials are defined as in \cite{kp1,kp2} by
\begin{eqnarray}\label{kravtchouk}
P_{N,k}^{(u,v)}(x)=\[{N\choose
k}(uv)^{k}\]^{-1/2}\sum_{\nu=0}^{k}
(-1)^{k-\nu}{x\choose \nu}{N-x\choose
k-\nu}u^{k-\nu}v^{\nu}\qquad
\(\begin{array}{l}0\le k \le N\cr
u+v=1\end{array}\)\,.
\end{eqnarray}
These functions form a complete system of orthogonal polynomials of integers
$x$ with $0\le x\le N$. They have quite diverse applications ranging from the
theory of covering codes \cite{cohen} to the statistical mechanics of polymers
\cite{schulten}. The same functions appear also as a building block in our
periodic-orbit theory of Anderson localization on graphs \cite{SS00}.
Unfortunately, we were not able to reduce the above expression any further by
using the known sum rules and asymptotic representations for Kravtchouk
polynomials.  The main obstacle stems from the fact that in our case the three
numbers $N,k,x$ in the definition (\ref{kravtchouk}) are constrained by
$N=k+x-1$.

\begin{figure}[tb]
\centerline{\psfig{figure=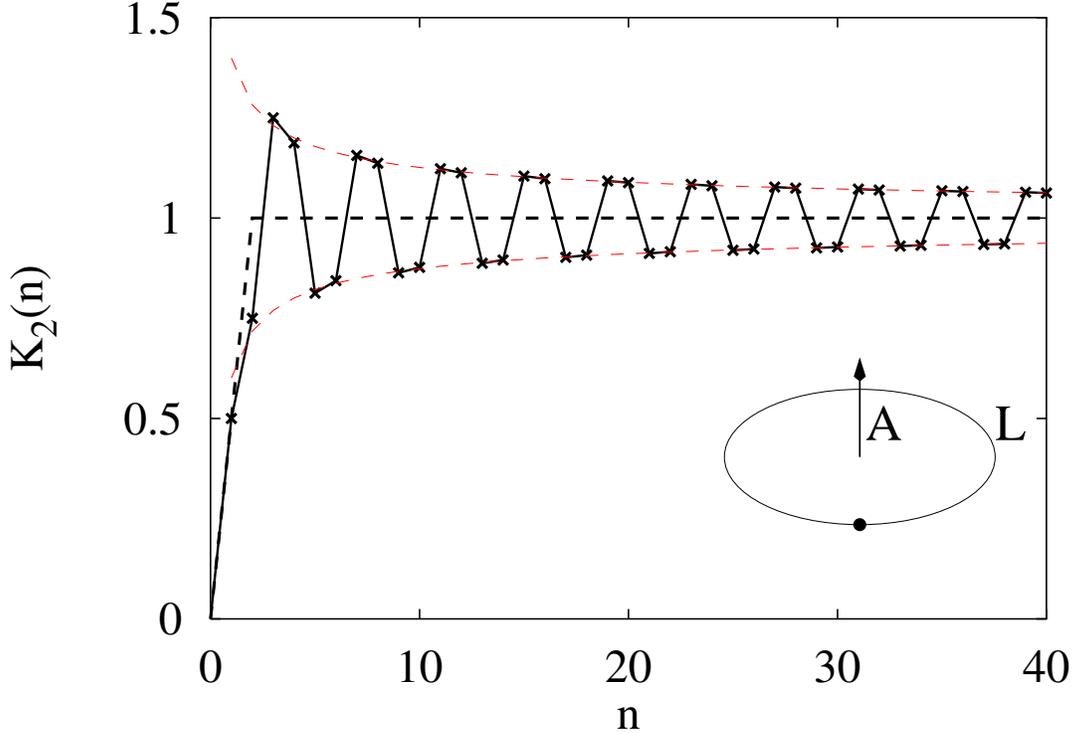,width=15cm,angle=270}
}
\caption{\label{tst} Form factor for the ring graph (see inset). The crosses
and the connecting heavy full line show the two equivalent exact results
(\protect\ref{K2PI4_UZY}) and (\protect\ref{k2pi4}) for $\eta=\pi/4$. The thin
dashed lines represent the approximation (\protect\ref{k2pi4_uzy_app}). The
heavy dashed line exhibits the form factor of a CUE ensemble of $2\times 2$
random matrices (\protect\ref{2-starK(n)CUE}), which can be obtained by an
appropriate averaging over $\eta$ (see text).}
\end{figure}

We will now consider the special case $\eta=\pi/4$ for which we obtained in
the previous subsection the solution (\ref{K2PI4_UZY}). The result can be
expressed in terms of Kravtchouk polynomials with $u=v=1/2$ which is also the
most important case for the applications mentioned above.  We adopt the common
practice to omit the superscript $(u,v)$ in this special case and find
\begin{eqnarray}\label{k2pi4}
K_2(n;\pi/4)&=&
{1\over
2^{n}}+{1\/2^{n+1}}\sum_{q=1}^{n-1}
{n-1\choose n-q}\[{n\over q}P_{n-1,n-q}(q)\right]^2\,.
\end{eqnarray}
It is convenient to introduce at this point the symbol $\nq(s,t)$ defined in
Eq.~(\ref{nq}). It allows to rewrite (\ref{k2pi4}) with the help of some
standard transformations of binomial coefficients as
\begin{eqnarray}\label{K2PI4N}
K_2(n;\pi/4)&=&{1\over 2^{n}}+{1\/2^{n+1}}\sum_{q=1}^{n-1}
\[{n\over q}\nq(q,n-q-1)\right]^2
\nonumber\\
&=&{1\over 2^{n}}+{1\over
2^{n+1}}\sum_{q=1}^{n-1}\[\nq(q,n-q)+(-1)^{n}\nq(n-q,q)\]^{2}
\end{eqnarray}
This expression is displayed in Fig.~\ref{tst} together with (\ref{K2PI4_UZY})
in order to illustrate the equivalence of the two results for the form factor.
%%%%%%%%%%%%%%%%%%%%%%%%%%%%%%%%%%%%%%%%%%%%%%%%%%%%%%%%%%%%%%%%%%%%%%%%%%%%%%%%
%%%%%%%%%%%%%%%%%%%%%%%%%%%%%%%%%%%%%%%%%%%%%%%%%%%%%%%%%%%%%%%%%%%%%%%%%%%%%%%%
An independent proof for the equivalence of (\ref{K2PI4_UZY}), (\ref{k2pi4})
can be given by comparing the generating functions of $K_2(n;\pi/4)$ in the
two representations \cite{gregory}. We define
\begin{eqnarray}\label{Gfun}
G(x)&=&\sum_{x=1}^{\infty}K_2(n;\pi/4)\,(2x)^{n}\qquad(|x|<1/2)
\end{eqnarray}
and find from (\ref{K2PI4_UZY})
\begin{eqnarray}\label{Gfun_uzy}
G(x)&=&{2x\/1-2x}-{1\over 2}+\sum_{m=0}^{\infty}{(-1)^{m}\/2}
{2m\choose m}x^{2m}(1-2x)
\nonumber\\
&=&{1\over 2}{1-2x\over \sqrt{1+4x^2}}-{1\over 2}{1-6x\/1-2x}\,.
\end{eqnarray}
On the other hand we have from (\ref{K2PI4N})
\begin{eqnarray}
G(x)={x\over 1-x}+G_{1}(x)+G_{2}(-x)
\end{eqnarray}
with $G_{1,2}(x)$ defined in the introduction.  A convenient starting point to
obtain the r.h.s. of (\ref{g1def}), (\ref{g2def}) is the integral
representation
\begin{eqnarray}\label{irep}
\nq(s,t)=-{(-1)^{t}\over 2\pi\i}\oint{\rm d}z\,(1+z^{-1})^{t}(1-z)^{s-1}\,,
\end{eqnarray}
where the contour encircles the origin. With the help of (\ref{irep}) we find
\begin{eqnarray}\label{g}
g(x,y)&=&\sum_{s,t=1}^{\infty}\nq(s,t)\,x^{s}\,y^{t}
\nonumber\\
&=&
-{1\over 2\pi\i}\sum_{s,t=1}^{\infty}
\oint{\rm d}z\,\sum_{s,t=1}^{\infty}(1+z^{-1})^{t}(1-z)^{s-1}\,x^{s}\,(-y)^{t}
\nonumber\\
&=&
{xy\over 2\pi\i}\sum_{s,t=0}^{\infty}\oint{\rm d}z\,
{1\over 1-x(1-z)}\,{1+z\over z+y(1+z)}
\nonumber\\
&=&
{xy\over (1+y)(1-x+y-2xy)}\qquad(|x|,|y|<1/\sqrt{2})\,,
\end{eqnarray}
which was already stated in (\ref{gdef}).  The contour
$|1+z^{-1}|=|1-z|=\sqrt{2}$ has been chosen such that both geometric series
converge everywhere on it. Now we have {\small
\begin{eqnarray}
G_{1}(x^2)&=&{1\over (2\pi\i)^{2}}\oint{{\rm
d}z\,{\rm
d}z'\/zz'}
\sum_{s,t=1}^{\infty}\sum_{s',t'=1}^{\infty}\nq(s,t)\,\nq(s',t')\,
(x
\,z)^{s}(x/z)^{s'}(x\,z')^{t}(x/z')^{t'}
\nonumber\\
&=&
{x^{4}\over
(2\pi\i)^{2}}\oint{{\rm d}z\,{\rm d}z'}\,
{1\over
(1+xz')(1+x[z'-z]-2x^2zz')}{z'\over
(z'+x)(zz'+x[z-z']-2x^2)}\,,
\end{eqnarray}
}
where $|x|<1/\sqrt{2}$ and the contour for $z,z'$ is the unit circle. We
perform the double integral using the residua inside the contour and obtain
(\ref{g1def}) and in complete analogy also (\ref{g2def}) such that finally
\begin{equation}\label{GfunPO}
G(x)={x\/1-x}+{x\over 2x-1}\({1\over\sqrt{4x^2+1}}-{1\/1-x}\)+
{1\over 2}{4x^2-2x+1\/(1-2x)\sqrt{4x^2+1}}-{1\over 2}\,.
\end{equation}
The proof is completed by a straight forward verification of the equivalence
between the rational functions (\ref{Gfun_uzy}) and (\ref{GfunPO}).

The identities (\ref{ci2a}), (\ref{ci2b}) follow now by separating even and odd
powers of $n$ in (\ref{K2PI4_UZY}) and (\ref{k2pi4}). In terms of
Kravtchouk polynomials these identities can be written as
\begin{equation}\label{ci2ak}
\sum_{q=1}^{2m-1}{2m-1\choose 2m-q}\[{2m\over
q}P_{2m-1,2m-q}(q)\]^{2}
=2^{2m+1}+(-1)^{m}{2m\choose m}-2
\end{equation}
and
\begin{equation}\label{ci2bk}
\sum_{q=1}^{2m}{2m\choose 2m+1-q}\[{2m+1\over q}P_{2m,2m+1-q}(q)\]^{2}
=2^{2m+2}-2\,(-1)^{m}{2m\choose m}-2\,.
\end{equation}
Finally we will derive the CUE result (\ref{2-starK(n)CUE}) for the ensemble
of graphs defined in the previous subsection starting from the periodic-orbit
expansion (\ref{K2eta}). We find
\begin{eqnarray}
K_{2}(n)&=&\int_{0}^{\pi/2}{\rm d}\mu(\eta)
K_2(n;\eta)\,.
\end{eqnarray}
Inserting (\ref{K2eta}), expanding into a double sum and using
\begin{equation}
\int_{0}^{\pi/2}{\rm d}\eta
\sin^{2(\nu+\nu')+1}\!\eta\cos^{2(n-\nu-\nu')+1}\!\eta= {1\over
2(n+1)}{n\choose \nu+\nu'}^{-1}
\end{equation}
we get
\begin{eqnarray}\label{intermediate}
K_{2}(n)&=&{1\/n+1}+
\\\nonumber
&&+{n^2\over 4(n+1)}\sum_{q=1}^{n-1}
\sum_{\nu,\nu'}
{(-1)^{\nu+\nu'}\/\nu\nu'}{n\choose
\nu+\nu'}^{-1}
{q-1\choose \nu-1}{n-q-1\choose \nu-1}{q-1\choose
\nu'-1}{n-q-1\choose
\nu'-1}\,.
\end{eqnarray}
Comparing this to the equivalent result (\ref{2-starK(n)CUE}) we were led to
the identity (\ref{ci3}) involving a multiple sum over binomial
coefficients. In this case, an independent computer-generated proof was found
\cite{akalu}, which is based on the recursion relation
\begin{equation}\label{recursion}
q^2F_{\nu,\nu'}(n,q)-(n-q-1)^{2}F_{\nu,\nu'}(n,q+1)+(n-1)(n-2q-1)
F_{\nu,\nu'}(n+1,q+1)=0\,.
\end{equation}
This recursion relation was obtained with the help of a Mathematica routine
\cite{multisum}, but it can be checked manually in a straight forward
calculation.  By summing (\ref{recursion}) over the indices $\nu,\nu'$, the
same recursion relation is shown to be valid for $S(n,q)$ \cite{multisum,A=B}
and the proof is completed by demonstrating the validity of (\ref{ci3}) for a
few initial values. Having proven (\ref{ci3}) we can use it to perform the
summation over $\nu,\nu'$ in (\ref{intermediate}) and find
\begin{eqnarray}
K_{2}(n)={1\/n+1}+\sum_{q=1}^{n-1}{n\/n^2-1}={1\/n+1}+{n\over
n+1}(1-\delta_{n,1})\,,
\end{eqnarray}
which is now obviously equivalent to the random-matrix form factor
(\ref{2-starK(n)CUE}).
%%%%%%%%%%%%%%%%%%%%%%%%%%%%%%%%%%%%%%%%%%%%%%%%%%%%%%%%%%%%%%%%%%%%
\subsection {\bf Trace identities}\label{trace}
Let $S$ be an arbitrary unitary matrix with a non degenerate spectrum and $s_n
= {\rm tr} S^n$. Then
\begin {eqnarray}
\lim_{\epsilon \rightarrow 0} \ \epsilon \ \sum _{n=n_0}^{\infty}
s_n^{*}\,s_{n+\nu}\,{\rm e}^{-n\epsilon}=s_{\nu} \ ,
\label {eq:identity1}
\end{eqnarray}
for arbitrary integers $n_0$ and $\nu$ \cite{US00}.

We shall now apply this identity to the ring graph with $\eta=\pi/4$
in order to prove (\ref{traceid}). Again, the traces of $S^{n}$ can be
expanded in periodic orbits which are grouped into $n+1$ families with equal
phases $\Phi(n,q) =q\phi_1+(n-q)\phi_2$ ($ 0\le q\le n$). Thus, one can write
\begin{equation}
 s_n(k) = \sum_{q=0}^n A(n,q)\,{\rm e}^{\i \Phi(n,q)}\,,
\label{eq:trace2star}
\end{equation}
where $A(n,q)$ is the coherent sum of all amplitudes of PO's in the
corresponding set. $A(n,q)$ can be expressed in terms of Kravtchouk
Polynomials as
\begin{equation}
A(n,q) =  {1\over\sqrt{2^{n}}} \left \{
\begin{array}{ll}
  1&
 {\rm for}\ \ \ q=0 \ {\rm or}\ n  \nonumber \\
  (-1)^{n+q} (n/q){n-1\choose n-q}^{1/2} P_{n-1,n-q}(q)\;\;\;&
  {\rm for} \ \ \ 0<q<n 
\end{array}
\right .
\end{equation}
(compare Eq.~(\ref{K2eta_K})). This is equivalent to (\ref{adef}).
Substituting (\ref {eq:trace2star}) into the trace identity (\ref
{eq:identity1}), we get for arbitrary integers $\nu$ and $n_0$
\begin {eqnarray}
\lim_{\epsilon \rightarrow 0} \ \epsilon &\sum\limits_{n=n_0}^{\infty}&
{\rm e}^{-n\epsilon} \sum _{q=0}^n \sum_{p=0}^{n+\nu}
A(n+\nu ,p) A(n,q)\,{\rm e}^{\i\left[(p-q)\phi_1 +(\nu-(p-q))\phi_2\right ]} 
\nonumber \\
 & =&\sum_{\kappa=0}^{\nu}  A(\nu,\kappa)\,{\rm e}^{\i \left [\kappa \phi_1
+(\nu-\kappa)\phi_2\right ]} \ .
\label{eq:identitykr}
\end{eqnarray}
This is valid for arbitrary phases $\phi_1$ and $\phi_2$ and therefore the
coefficients of the phase factors ${\rm e}^{\i \Phi(\nu,\kappa)}$ on both
sides are equal. Eq.~(\ref{traceid}) follows.

%%%%%%%%%%%%%%%%%%%%%%%%%%%%%%%%%%%%%%%%%%%%%%%%%%%%%%%%%%%%%%%%%%%%%%
\section{\bf Conclusions}
We have shown how within periodic-orbit theory the problem of finding the form
factor (the spectral two-point correlation function) for a quantum graph can
be reduced exactly to a well-defined combinatorial problem. Even for the very
simple graph model that we considered in the last section the combinatorial
problems involved were highly non-trivial. In fact we encountered previously
unknown identities which we could not have obtained if it were not for the
independent method of computing the form factor directly from the spectrum.
However, since the pioneering work documented in \cite{A=B} the investigation
of sums of the type we encountered in this paper is a rapidly developing
subject, and it can be expected that finding identities like (\ref{ci3}), 
(\ref{ci2a}) and (\ref{ci2b}) will shortly be a matter of computer power.

Numerical simulations in which the form factor was computed for fully
connected graphs ($v_i=V-1\ \forall i$) and, e.~g., Neumann boundary
conditions indicate that the spectral correlations of $S$ match very well
Dyson's predictions for the circular ensembles COE or CUE, respectively
\cite{KS97}. The agreement is improved as $V$ increases.  We conjecture that
this might be rigorously substantiated by asymptotic combinatorial theory. A
first step towards this goal was taken in the present paper.
%%%%%%%%%%%%%%%%%%%%%%%%%%%%%%%%%%%%%%%%%%%%%%%%%%%%%%%%%%%%%%%%%%%%%%
\section{\bf Acknowledgements}
This research was supported by the Minerva Center for Physics of Nonlinear
Systems, and by a grant from the Israel Science Foundation.  We thank Uri
Gavish for introducing us to the combinatorial literature, and Brendan McKay
and Herbert Wilf for their interest and support. We are indebted to Gregory
Berkolaiko for his idea concerning the proof of (\ref{ci2a}) and (\ref{ci2b}),
and to Akalu Tefera for his kind help in obtaining a computer-aided proof of
(\ref{ci3}).
%%%%%%%%%%%%%%%%%%%%%%%%%%%%%%%%%%%%%%%%%%%%%%%%%%%%%%%%%%%%%%%%%%%%%%

\end{document}